# H I Self-absorption toward the Cygnus X North: From Atomic Filament to Molecular Filament

Chong Li[1,2], Keping Qiu[1,2], Di Li[3,4,5], Hongchi Wang[6,7], Yue Cao[1,2], Junhao Liu[1,2], Yuehui Ma[6,7], and Chenglin Yang[8]
[1] School of Astronomy and Space Science, Nanjing University, 163 Xianlin Avenue, Nanjing 210023, People's Republic of China; kpqiu@nju.edu.cn
[2] Key Laboratory of Modern Astronomy and Astrophysics (Nanjing University), Ministry of Education, Nanjing 210023, People's Republic of China
[3] National Astronomical Observatories, Chinese Academy of Sciences, Beijing 100101, People's Republic of China
[4] Research Center for Intelligent Computing Platforms, Zhejiang Laboratory, Hangzhou 311100, People's Republic of China
[5] NAOC-UKZN Computational Astrophysics Centre, University of KwaZulu-Natal, Durban 4000, South Africa
[6] Purple Mountain Observatory and Key Laboratory of Radio Astronomy, Chinese Academy of Sciences, 10 Yuanhua Road, Nanjing 210033, People's Republic of China
[7] School of Astronomy and Space Science, University of Science and Technology of China, 96 Jinzhai Road, Hefei 230026, People's Republic of China
[8] University of Chicago, 5801 S Ellis Ave, Chicago, IL 60637, USA
Received 2023 March 16; revised 2023 April 21; accepted 2023 April 24; published 2023 May 10

## Abstract

Using the H I self-absorption data from the Five-hundred-meter Aperture Spherical radio Telescope, we perform a study of the cold atomic gas in the Cygnus X North region. The most remarkable H I cloud is characterized by a filamentary structure, associated in space and in velocity with the principal molecular filament in the Cygnus X North region. We investigate the transition from atomic filament to molecular filament. We find that the H II regions Cygnus OB2 and G081.920+00.138 play a critical role in compressing and shaping the atomic Cygnus X North filament, where the molecular filament subsequently forms. The cold H I in the DR21 filament has a much larger column density ($N$(H I) $\sim 1 \times 10^{20}$ cm$^{-2}$) than the theoretical value of the residual atomic gas ($\sim 1 \times 10^{19}$ cm$^{-2}$), suggesting that the H I-to-H$_2$ transition is still in progress. The timescale of the H I-to-H$_2$ transition is estimated to be $3 \times 10^5$ yr, which approximates the ages of massive protostars in the Cygnus X North region. This implies that the formation of molecular clouds and massive stars may occur almost simultaneously in the DR21 filament, in accord with a picture of rapid and dynamic cloud evolution.

*Unified Astronomy Thesaurus concepts:* Interstellar atomic gas (833); Interstellar filaments (842); Molecular clouds (1072); Star forming regions (1565)

## 1. Introduction

In the last decade, Herschel observations have revealed the ubiquitous presence of filaments of molecular gas (André et al. 2010, 2014) and their close relationship with star formation (e.g., Schneider et al. 2012). The molecular filaments have been observed in several tracers, ranging from extinction maps at optical and near-infrared bands (e.g., Jackson et al. 2010; Kainulainen et al. 2013) to far-infrared/submillimeter dust emission maps (e.g., Men'shchikov et al. 2010; Schneider et al. 2010) and CO maps (e.g., Yuan et al. 2021; Guo et al. 2022). Molecular filaments can be produced by dynamical models with turbulence (Vazquez-Semadeni 1994; Padoan et al. 2001), converging flows (Elmegreen 1993; Vázquez-Semadeni et al. 2006; Heitsch & Hartmann 2008; Clark et al. 2012), instabilities in self-gravitating sheets (Nagai et al. 1998), or other processes that compress the gas to an overdense interface (Padoan et al. 2001). However, the dominant mechanism of molecular filament formation is still in debate.

Recent H I observations present a picture that the cold neutral interstellar medium (ISM) is mainly distributed in filamentary structures (Kalberla et al. 2016, 2020; Soler et al. 2020). These H I filaments are found to be preferentially aligned along the magnetic field and associated with dust emission. Supporting evidence for low temperatures at the position of H I filaments was recently reported from Na I absorption measurements (Peek & Clark 2019). Although both atomic and molecular gases show rich filamentary structures, whether atomic filaments can evolve into molecular filaments remains open.

Both simulations (e.g., Wolfire et al. 2003) and observations (e.g., Heiles & Troland 2003a, 2003b) have revealed that the atomic gas exists in two stable phases, i.e., the so-called cold neutral medium (CNM; <300 K, $n \sim 20$ cm$^{-3}$) and warm neutral medium (WNM; 5000–10,000 K, $n \sim 0.2$ cm$^{-3}$) (Kalberla & Kerp 2009; Girichidis et al. 2020). Some observations also suggest the presence of a thermally unstable WNM (e.g., Begum et al. 2010). The H$_2$ transition rate from diffuse atomic hydrogen is proportional to the volume density of H I (Li & Goldsmith 2003; Goldsmith & Li 2005; Goldsmith et al. 2007). Therefore, while there is a higher percentage of atomic hydrogen in the warm phase compared to the cold component (Kalberla & Haud 2018), the CNM is the key component in the conversion from atomic hydrogen to its molecular phase. However, due to the coexistence of multiple phases of neutral atomic hydrogen, it is difficult to characterize from observations the properties of CNM from which the molecular hydrogen forms. Such a difficulty can be overcome by observing the H I self-absorption (HISA), as it only traces the cold atomic gas (e.g., Gibson et al. 2005a, 2005b). In particular, the H I narrow (less than that of CO) line-width self-absorption (HINSA; Li & Goldsmith 2003) traces the coldest H I within molecular clouds.

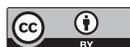

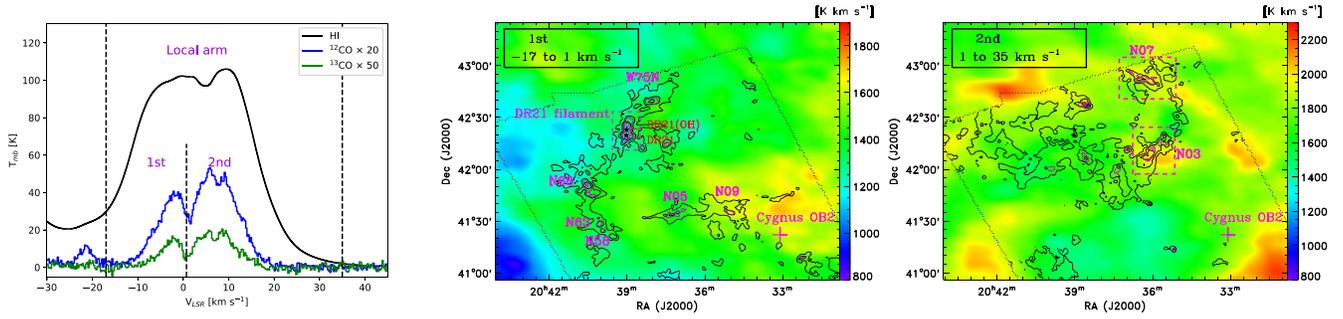

**Figure 1.** Left: average spectra of the 3°.3 × 2°.5 region of Cygnus X North with the black, blue, and green lines indicating the H I, $^{12}$CO, and $^{13}$CO emissions, respectively. Middle: line-integrated maps of H I (colors in the backgrounds), $^{12}$CO (black contours), and $^{13}$CO (purple contours) emissions of the first velocity component. The minimal level and the interval of the black contours are 30 K km s$^{-1}$ and 45 K km s$^{-1}$, respectively. The minimal level and the interval of the purple contours are both 21 K km s$^{-1}$. The dashed lines indicate the coverage of the $^{12}$CO and $^{13}$CO data. Right: line-integrated maps of H I, $^{12}$CO, and $^{13}$CO emissions of the second velocity component. The minimal level and the interval of the black contours are 50 K km s$^{-1}$. The minimal level and the interval of the red contours are 24 K km s$^{-1}$.

The Cygnus X region has already been shown to be an excellent laboratory for studying high-mass star formation. The large-scale molecular clouds (>10 pc; Schneider et al. 2006; Rygl et al. 2012), medium-scale dynamic structure properties (1–10 pc; Motte et al. 2007; Schneider et al. 2010), and dense cores on 0.01–0.1 pc scales (Cao et al. 2019, 2021) in the region have been investigated. The principal molecular cloud in the Cygnus X North region consists of three entangled fibers in the north (Cao et al. 2022) and a coherent cold molecular filament in the south (Hu et al. 2021). Although the properties of molecular clouds (e.g., Schneider et al. 2010), dust (e.g., Cao et al. 2019), ionized gas (e.g., Wang et al. 2022), and star formation processes (e.g., Motte et al. 2018) have been studied around the Cygnus X North region, the properties of atomic gas did not receive much attention in previous surveys.

In this work we present a large-scale (3°.3 × 2°.5) H I observation toward the Cygnus X North region, which is part of the comprehensive surveys of the Cygnus X complex (CENSUS; Cao et al. 2019; Hu et al. 2021; Cao et al. 2021, 2022; Wang et al. 2022). We focus on the HISAs. Evidence for the transition from an atomic filament to a molecular filament is presented. The observation strategy and data reduction details are described in Section 2, and the results are presented in Section 3. We discuss our results in Section 4 and present the summary in Section 5.

## 2. Observations and Data Reduction

### 2.1. FAST H I Data

Using FAST, we observed H I 1.420 GHz line emission toward the Cygnus X North region on 2019 August 24. The observations cover the sky region of 307°.7 < α < 311°.0 and 40°.9 < δ < 43°.4 (3°.3 × 2°.5). The FAST L-band Array of 19 feed horns (FLAN) was used as the front end (Li et al. 2018b). The narrow ROACH back-end possesses a bandwidth of 31.125 MHz and contains 65,536 channels, resulting in a spectral resolution of 0.476 kHz and a velocity resolution of 0.1 km s$^{-1}$ at 1.420 GHz (Jiang et al. 2020). The sky region was scanned along the R.A. in multibeam on-the-fly (OTF) mode with a scanning rate of 15″ per second and a dump time of 1 s. The rotating angle between the 19-beam focal plan array and direction of decl. was set to 23°.4 for smooth super-Nyquist sampling. The pointing of the telescope has an rms accuracy of 7″.9.

For intensity calibration, a 1.1 K noise from the diode was injected with a period of 2 s during observations, which is synchronized with the sampling rate. Based on an absolute measurement of the noise dipole and a factor derived by the difference between the noise ON and OFF data for each beam, the units of observed data were calibrated into the antenna temperature $T_A$ in kelvins (Jiang et al. 2020). We further compared our data with the HI4PI data (HI4PI Collaboration et al. 2016), which has been corrected for stray radiation. We attribute the integrated intensity (−17–35 km s$^{-1}$) difference between our data and the HI4PI data to the stray radiation and consequently scaled our data.

Using a Gaussian smoothing kernel, the raw data are regridded and converted to an FITS data cube. The beamwidth is 3′ and the pixel size of the FITS data cube is 1′ × 1′. The typical system temperature during the observation is approximately 20 K, and the rms sensitivity of our observation is estimated to be around 70 mK per channel.

### 2.2. Nobeyama CO Data

The $^{12}$CO and $^{13}$CO data used in this work were obtained from the Nobeyama 45 m Cygnus X CO Survey (Yamagishi et al. 2018; Takekoshi et al. 2019). The simultaneous observations of $^{12}$CO (J = 1-0) and $^{13}$CO (J = 1-0) cover both Cygnus X North and South as well as the Cygnus OB2 association. The OTF scan parameters and calibration method are the same as those of the FUGIN Galactic plane survey (Umemoto et al. 2017). The RX angle, scan spacing, scan length, scan speed, and sampling time are 9″.46, 8″.5, 3600″, 100″ s$^{-1}$, and 40 ms, respectively. The typical system temperatures are 350 K for $^{12}$CO and 150 K for $^{13}$CO. The effective angular resolution is 46″, and the velocity resolution is 0.25 km s$^{-1}$. The noise levels are estimated to be approximately 0.88 K for the $^{12}$CO J = 1-0 emission and 0.36 K for the $^{13}$CO J = 1-0 emission in the $T_{mb}$ scale. It should be noted that the $^{12}$CO and $^{13}$CO data do not cover the whole region of our observations. However, the vast majority of the molecular clouds in the Cygnus X North region have been included in their data.

### 2.3. Overall Distributions of the Atomic and Molecular Gas in Cygnus X North

The average spectra and the integrated intensity maps of the H I, $^{12}$CO, and $^{13}$CO in the Cygnus X North region are displayed in Figure 1. The Cygnus X North complex has a velocity range from $-17$ to 35 km s$^{-1}$. The $^{12}$CO and $^{13}$CO spectra show that the molecular clouds can mainly be attributed to two velocity components, with velocity ranges from $-17$ km s$^{-1}$ to 1 km s$^{-1}$ and from 1 km s$^{-1}$ to 35 km s$^{-1}$, respectively. The clouds with velocities from $-3$ to $-17$ km s$^{-1}$ are located in a region between the Local arm and Perseus arm (C. Li et al., 2023, in preparation), which is not the focus of this work. The major emission of the first velocity component comes from the dense regions N58, N63, N68, DR21 filament, W75N, N05, and N09 (Motte et al. 2007; Cao et al. 2019). The peak $^{13}$CO emission of the Cygnus X North molecular cloud is located in the DR21 filament, including the high-mass star-forming sites DR21 and DR21(OH). Due to the diffuse and multiphase nature of the neutral gas, the distribution of the H I-integrated intensity map does not coincide with that of the molecular cloud. For the second velocity component, the molecular clouds show relatively smooth distributions, with the densest clouds located in the N03 and N07 regions. Velocity channel maps of H I and $^{12}$CO emissions are shown in Appendix A. The most remarkable feature in the Cygnus X North molecular cloud is a filamentary structure oriented in the north–south direction. The second velocity component has diffuse morphology, located in a cavity of the H I emission.

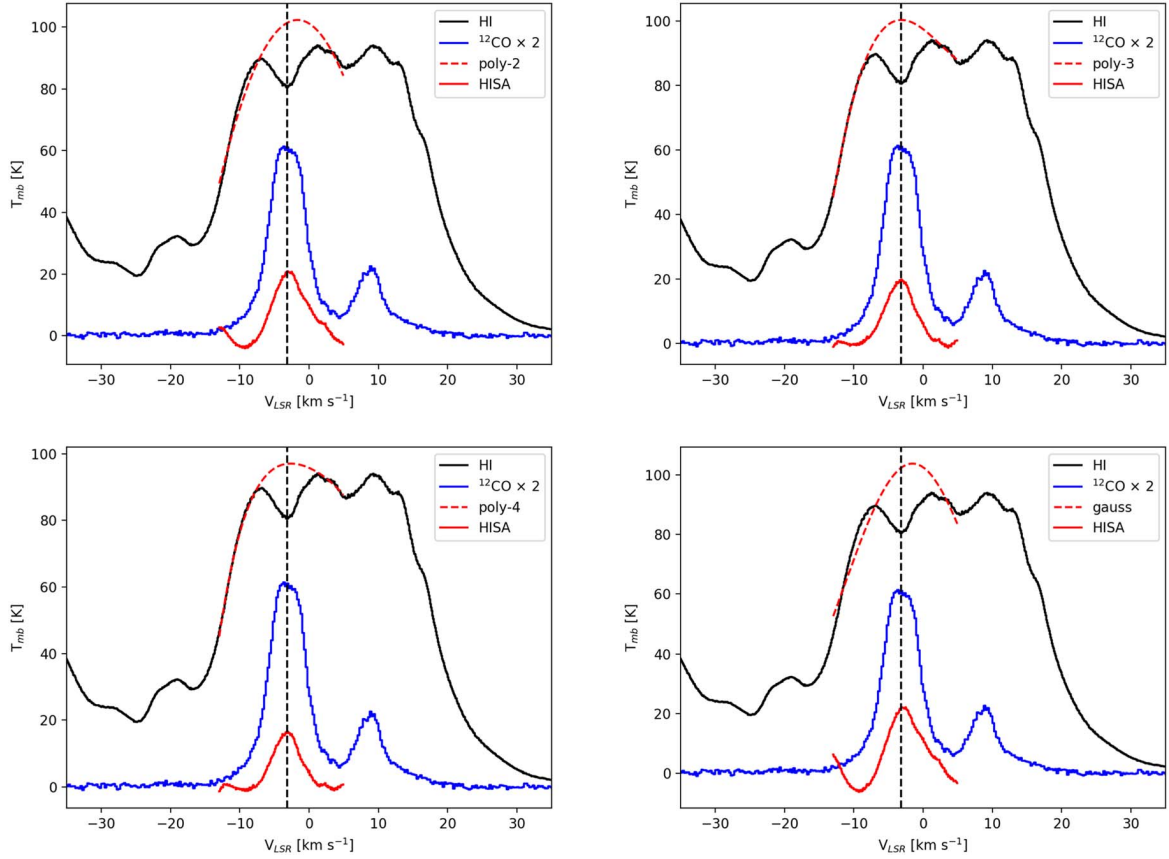

**Figure 2.** Average H I ($T_{on}$), HISA, and $^{12}$CO spectra over $5' \times 5'$ centered on the DR21 filament region. The dashed red lines show the fitted curves ($T_{off}$) to the absorption-free channels of the H I spectrum using different methods. The HISA curve (solid red line) is estimated by subtracting the H I spectrum from the fitted curve.

### 2.4. Extraction of the H I Self-absorption Profile

To extract the HI self-absorption profile, we test several approaches. In Figure 2 we show the HI self-absorption features of the DR21 filament region extracted using the second-order, third-order, and fourth-order polynomials and Gaussian fits, respectively. The third-order and fourth-order polynomial approaches exhibit good performance in deriving flat baselines. We further compared the velocities of peak emissions in extracted HI self-absorption spectra with those in $^{12}$CO emission in the Cygnus X North region. The third-order polynomial fit shows the best performance in matching the H I peak velocities with those of $^{12}$CO (see Appendix B). We adopt the third-order polynomial fit in the following analysis. The HI self-absorption has an FWHM line width (4.8 km s$^{-1}$) less than that of $^{12}$CO (6.3 km s$^{-1}$), indicating that it is in nature a H I narrow self-absorption (HINSA). Considering that the line widths here are larger than the typical value (1∼2 km s$^{-1}$) in Li & Goldsmith (2003), which could be caused by the fact that the corresponding molecular clouds are more active and warmer, we still term the HI self-absorption as HISA in the following analyses. The $^{12}$CO emission in the DR21 filament has a highest brightness temperature of ∼40 K, which approximates the excitation temperature.

### 3. Results

#### 3.1. H I Self-absorption toward the Cygnus X North Filament

The $^{12}$CO emission shows that the first velocity component of the molecular gas exhibits a filamentary structure (the left panel of Figure 3), which we refer to as the Cygnus X North filament hereafter. The Cygnus X North filament consists of the dense regions ($N_{H_2} \geqslant 3.5 \times 10^{22}$ cm$^{-2}$) N58, N63, N68, DR21 filament (DR21 and DR21(OH)), and W75N from south to north (Motte et al. 2007; Cao et al. 2019). The length of the Cygnus X North filament is 46 pc, which means it is a giant molecular filament. The arcuate shape of the Cygnus X North filament coincides well with the east protruding edge of the integrated H I emission (the left panel of Figure 3), implying the presence of HISA (Li & Goldsmith 2003; Kavars et al. 2005). In the position–velocity map (the middle panel of Figure 3), the CO emission exactly follows the central self-absorption of the H I emission. Thus the Cygnus X North filament correlates well in both position and velocity with the HISA distribution revealed by the FAST observations, providing strong evidence that we are witnessing a remarkable molecular filament forming out of an atomic filament.

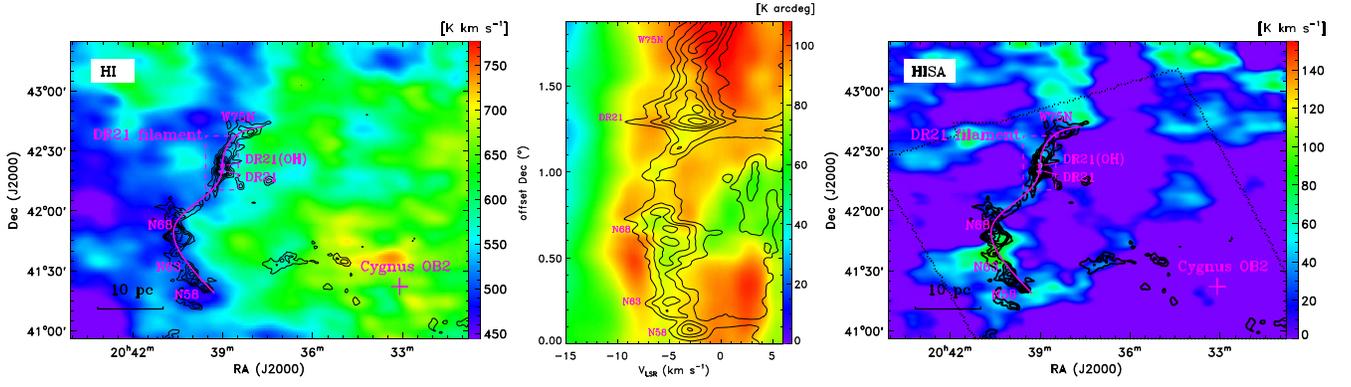

**Figure 3.** Left: $^{12}$CO integrated intensity overlaid on the line-integrated H I emission. The purple line indicates distribution of the Cygnus X North filament. The intensities are integrated over a velocity range from $-8$ to $-2$ km s$^{-1}$. The minimal level and the interval of the overlaid contours are $0.2\times$ peak and $0.15\times$ peak of $^{12}$CO brightness. Middle: position–velocity map of H I (color) and $^{12}$CO (contours) emission along the purple line in the left panel. Right: map of HISA intensity integrated from $-8$ to $-2$ km s$^{-1}$.

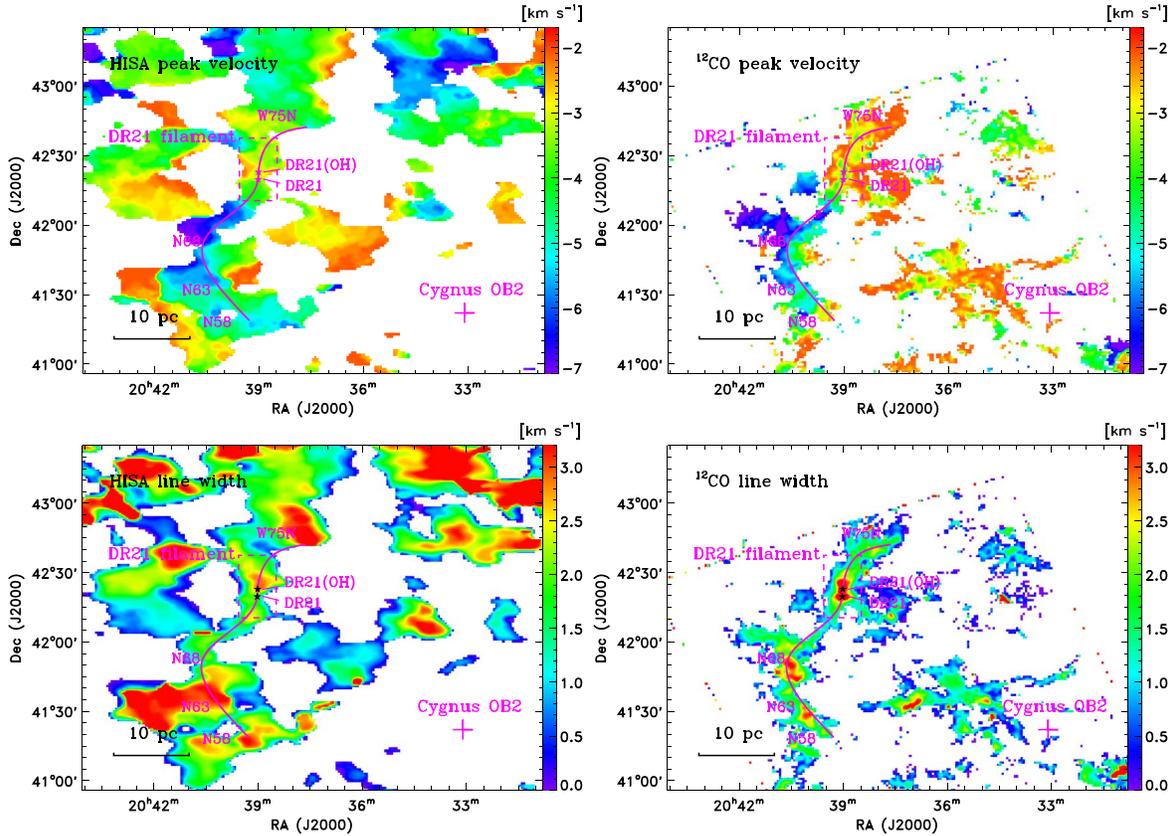

**Figure 4.** Top row: peak velocity of the HISA data (left panel) and the $^{12}$CO data (right panel), respectively. The HISA are extracted by using a third-order polynomial to estimate the background emissions. Bottom row: FWHM of the HISA emission (left panel) and the $^{12}$CO feature (right panel), respectively.

To further check the HISA distribution, we extract the HISA profile of the Cygnus X North region (see Section 2.4). We display the velocity-integrated map of the HISA features in the right panel of Figure 3, which is the first large-scale spatial distribution of HISA made with the FAST. The HISA structure shows a filamentary morphology corresponding to the Cygnus X North molecular filament. The peak velocity of $^{12}$CO and that of HISA are presented in the top panels of Figure 4. The velocity distribution of HISA is generally consistent with that of the $^{12}$CO emission. The velocities of the DR21 filament and W75N are relatively redshifted, and the velocity of N68 is blueshifted. The line widths of both $^{12}$CO and HISA are shown in the bottom panels of Figure 4. The line width for the $^{12}$CO emission shows relatively high values of more than 3 km s$^{-1}$ for the DR21 filament, N68, and N63 regions, most likely attributed to active star formation feedback.

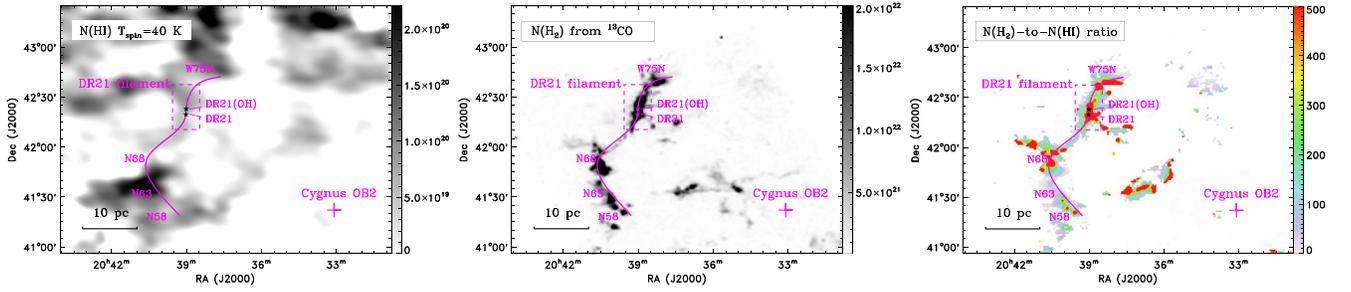

**Figure 5.** Left: column density map of the H I cloud derived from HISA, assuming the spin temperature $T_{HISA} = 40$ K and P = 0.7. Middle: column density map of the molecular cloud derived from $^{13}$CO emission. Right: ratio of the H$_2$-to-H I column densities.

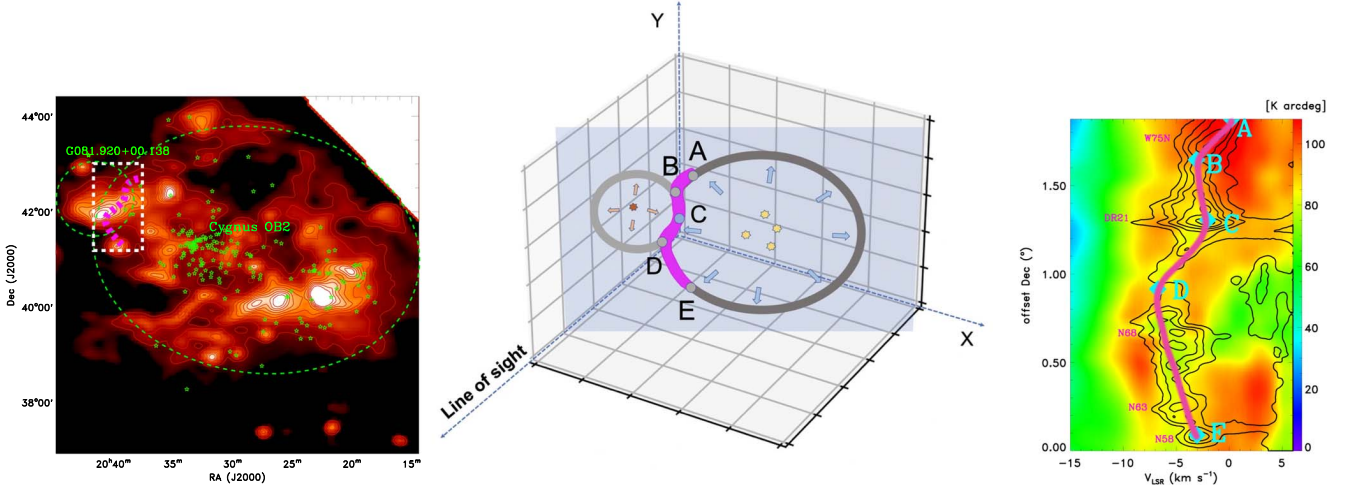

**Figure 6.** Left: 1.4 GHz radio continuum emission from the Effelsberg Galactic plane survey (Reich et al. 1990, 1997). The circle sizes approximate the radius of the H II regions from Anderson et al. (2014). The purple line indicates the distribution of the Cygnus X North filament. The green stars indicate the O-type stars in Cygnus OB2 from Comerón & Pasquali (2012). Middle: the sketch map of H II regions Cygnus OB2 and G081.920+00.138 in the three-dimensional coordinate system. Right: position–velocity map of H I (colors in the backgrounds) and $^{12}$CO (contours) emission along the decl. in the left panel.

A comprehensive discussion of the radiative transfer of HISA is given by Gibson et al. (2000), Li & Goldsmith (2003), and Wang et al. (2020). Assuming the foreground and background clouds are optically thin, the spin temperature of a HISA cloud follows

$$T_{on-off} = (T_{HISA} - pT_{off} - T_{cont}) \times (1 - e^{-\tau_{HISA}}) \quad (1)$$

$$p = -\frac{T_{bg}(1 - e^{-\tau_{bg}})}{T_{off}}. \quad (2)$$

$T_{on}$ and $T_{off}$ are the H I brightness temperatures in the presence and absence of self-absorption, respectively (see the solid black and dashed red curves in Figure 2). Assuming optically thin emission, $p = 1$ implies that there is no foreground emission, and $p = 0.5$ means that the foreground and background emissions are equal. Assuming that the H I emission with velocity $> 1$ km s$^{-1}$ is from the foreground cloud and the emission with velocity $< -17$ km s$^{-1}$ is from the background, we estimated the parameter $p$ to be $\sim 0.7$. As shown in Equation (1), we cannot solve the optical depth and the spin temperature simultaneously. We assume that the HISA cloud has the same temperature as the molecular cloud, which is around 40 K from $^{12}$CO. Adopting $T_{cont}$ from the Effelsberg 21 cm radio continuum survey (Reich et al. 1990, 1997), we derive the optical depth of HISA in the Cygnus X North region (Appendix C) and take them to correct for the column density of cold atomic gas. The median optical depth is around 0.3, and the maximum optical depth is located in the N63 region, which is around 1.6. Using the methods in Wang et al. (2020) and Syed et al. (2020), we calculated the column densities of the cold atomic and molecular hydrogen (Figure 5). The median column density of the cold H I gas for the Cygnus X North filament is around $1 \times 10^{20}$ cm$^{-2}$. The N63 region has the largest column density, which is about $3.0 \times 10^{20}$ cm$^{-2}$. The median column density of the molecular gas is about $7.8 \times 10^{22}$ cm$^{-2}$. The Cygnus X North filament exhibits a H$_2$-to-H I column density ratio of $\sim 300$, which is comparable to the ratios for the nearby clouds found by Zuo et al. (2018; 50–500). The DR21, DR21(OH), W75N, and N68 show the largest H$_2$-to-H I ratio, which is about 500.

### 4. Discussion

#### 4.1. The Formation of the Cygnus X North Filament

H II regions play an important role in shaping their nearby clouds and regulating the evolution of ISM. Figure 6 shows the distributions of the 1.4 GHz radio continuum emission in the sky region. The OB star association Cygnus OB2 has received much attention and has been studied at all wavelengths with different spatial coverages since it possesses a high number of early spectral-type stars. Hosting around 100 O-type stars (Knödlseder 2000), it is one of the most massive associations in our Galaxy and produced the remarkable H II region. To the east of this dominant H II region, there is a relatively small H II region G081.920+00.138. The Cygnus X North filament is located at the interaction area of the two H II regions, pointing to the picture that clouds could be formed from the converging flows driven by nearby massive stars (Hartmann et al. 2001). Besides that, the bow shape of the Cygnus X North filament coincides well with the protruding edges of the interaction regions, which further suggests that the HISA cloud of the Cygnus X North filament is shaped by the interaction of the two H II regions.

The parallaxes of OB stars in the Cygnus OB2 from Gaia satellite data release 2 (DR2; Gaia Collaboration et al. 2018; Bailer-Jones et al. 2018) are around 0.6 mas, which corresponds to 1.6 ~1.7 kpc (Orellana et al. 2021). The OB stars in the H II region G081.920+00.138 have parallaxes of ~0.7 mas (1.4 kpc; Xu et al. 2018, 2021). In other words, Cygnus OB2 is located at a farther distance than G081.920+00.138 (middle panel in Figure 6). Without the influence of G081.920+00.138, the C position should be the nearest to us and have the most blueshifted velocity, with the B/D and A/E positions being increasingly farther away from us. However, the G081.920+00.138 H II region is also expanding and pushes the C position away from us, making its velocity a bit redshifted. Such a collision picture is consistent with the position–velocity map along the Cygnus X North filament (right panel in Figure 6), which corroborates the picture that the H II regions have influence on the atomic and molecular gas. The velocities of A, C, and E are different from those of B and D by about 5 km s$^{-1}$, corresponding to a velocity gradient of 0.5 km s$^{-1}$ pc$^{-1}$. Such a velocity gradient is usual in the gas around the H II regions (Dent et al. 2009; Gong et al. 2016; Li et al. 2018a). The median column density of the cold H I for the Cygnus X North filament is around $1 \times 10^{20}$ cm$^{-2}$, much larger than the typical value of H I filaments in our Galaxy (~$10^{19}$ cm$^{-2}$; Clark et al. 2014; Kalberla et al. 2016, 2020), which can be explained by the compression of the H II regions Cygnus OB2 and G081.920+00.138. Based on the above picture, the interaction of two expanding H II regions creates the relatively dense H I filament, from which the molecular filament forms.

### 4.2. Cloud Age Based on the H I-to-H$_2$ Transition Stage

Basically, atomic gas cools down to form molecules on the grain surface. The formation rate of H$_2$ molecules on the grain surfaces with a specified cross section and volume density is given by Hollenbach et al. (1971). However, it is difficult to model the production of H$_2$ given the wide range of dust size and the photodestruction by the interstellar radiation. In order to calculate the transition from atomic to molecular hydrogen, a more practical scenario for the formation of H$_2$ is given by Goldsmith & Li (2005). Assuming that all dust grains are spherical particles with different size, the modeling cloud is designed to be initially ($t = 0$) entirely in atomic form. The beginning of the H$_2$ formation is set to be that atomic gas contracts to be optically thick enough ($A_v \sim 0.5$, $N(\mathrm{H\,I}) \sim 1 \times 10^{21}$ cm$^{-2}$) to shield the photodestruction. Assuming the evolution of the cloud takes place at a constant density and the only destruction pathway for molecular hydrogen is cosmic-ray, the time dependence of the density

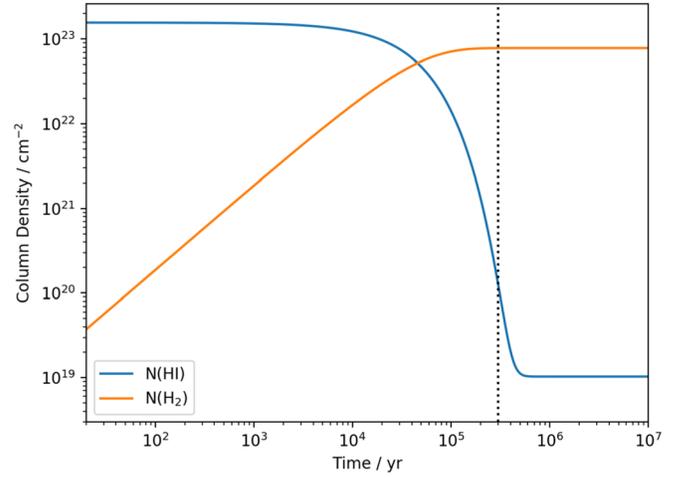

**Figure 7.** $N(\mathrm{H\,I})$ and $N(\mathrm{H}_2)$ as a function of time for the DR21 filament. The black dotted line indicates the calculated time from the $N(\mathrm{H\,I})$ and $N(\mathrm{H}_2)$ of the DR21 filament.

**Table 1**
Lifetime Estimates Related to Massive Star Formation in the Cygnus X North Region

|  | Lifetime (yr) | References |
| --- | --- | --- |
| Cloud age | $3 \times 10^5$ | this work |
| UCHII regions | $3 \times 10^5$ | Motte et al. (2018) |
| Massive protostars | $3 \times 10^4$ | Motte et al. (2007) |
| Outflow in DR21 | $1 \times 10^4$ | Zapata et al. (2013) |

of molecular hydrogen obeys

$$\frac{dn_{\mathrm{H}_2}}{dt} = k' n_{\mathrm{H\,I}} n_0 - \zeta_{\mathrm{H}_2} n_{\mathrm{H}_2}, \quad (3)$$

where $k' n_{\mathrm{H\,I}} n_0$ and $\zeta_{\mathrm{H}_2} n_{\mathrm{H}_2}$ are the formation rate and destruction rate of H$_2$, respectively. The factors $n_{\mathrm{HI}}$ and $n_{\mathrm{H}_2}$ are the atomic and molecular hydrogen density, respectively. The factor $n_0$ is the total proton density in atomic and molecular hydrogen, $n_0 = n_{\mathrm{HI}} + 2 n_{\mathrm{H}_2}$. The factor $k'$ depends on the maximum and minimum values of the grain size distribution, which is found to have a nominal value of $k' = 1.2 \times 10^{-17}$ cm$^3$ s$^{-1}$ (Goldsmith & Li 2005; Goldsmith et al. 2007). The factor $\zeta_{\mathrm{H}_2}$ is the cosmic-ray destruction rate, which is found to lie in the range of $10^{-18}$–$10^{-16}$ s$^{-1}$ in dense clouds (Caselli et al. 1998) and then has been more narrowly constrained to be $5 \times 10^{-17}$ s$^{-1}$ in other studies (Bergin et al. 1999; van der Tak & van Dishoeck 2000), which is consistent with the results of Doty et al. (2002) from detailed modeling of the source AFGL 2591. We focus on the densest region of the Cygnus X North filament, i.e., the DR21 filament. Applying Gaussian fitting to the H$_2$ column density along the DR21 filament, we derive the FWHM width to be 1.6 pc. In Figure 7 we show the cloud age based on Equation (3), taking $k' = 1.2 \times 10^{-17}$ cm$^3$ s$^{-1}$, $\zeta_{\mathrm{H}_2} = 5 \times 10^{-17}$ s$^{-1}$, $n_{\mathrm{H}_2} = N(\mathrm{H}_2)/1.6 \mathrm{pc}$, and $n_{\mathrm{HI}} = N(\mathrm{H\,I})/1.6$ pc. According to the current column densities of $N(\mathrm{H\,I})$ and $N(\mathrm{H}_2)$, the cloud age of the DR21 filament is calculated to be ~$3 \times 10^5$ yr, which approximates the typical age of the UCHII regions (Motte et al. 2018) and the dynamic lifetime of the subfilaments in DR21 filament (Schneider et al. 2010). The DR21 region in the filament is a very active massive star-forming region and hosts the most energetic young stellar object outflow in the Galaxy, which has a dynamical timescale of about $1 \times 10^4$ yr (Zapata et al. 2013). The massive protostars in the region have ages about $3 \times 10^4$ yr (Motte et al. 2007). Considering all the timescales above (see Table 1), we could speculate a very dynamic scenario in which the expanding H II regions compress the atomic gas, forming a dense and cold H I layer and promoting the HI-to-H$_2$ transition, creating a remarkable molecular filament. Within the molecular filament, massive star formation commences quickly, almost simultaneously, with the filament formation, as long as local high-density peaks are introduced by compressive and turbulent shocks. Such a scenario is in general consistent with a picture of dynamic molecular clouds where molecular clouds form and evolve rapidly and form stars quickly (e.g., Hartmann et al. 2001).

## 5. Summary

Using the FAST, we carried out a study of the cold atomic gas in the Cygnus X North region. The cold atomic hydrogen is traced through the detected self-absorption features (HISAs). By combining the molecular clouds traced by $^{12}$CO and $^{13}$CO emissions from the Nobeyama 45 m Cygnus X CO Survey, the properties of the cold H I gas related to the molecular clouds are presented. The main results are summarized as follows:

1. Both the atomic and the molecular clouds with velocities lying in the range from −8 to −2 km s$^{-1}$ exhibit filamentary structures. The shape, location, and velocity of the atomic filament coincide with those of the molecular filament, implying that the atomic filament is the birthplace of the molecular filament. Atomic filaments could be the direct progenitors of molecular filaments.

2. The H II regions, i.e., the Cygnus OB2 and G081.920 +00.138, are interacting to compress the gas and help to form the atomic Cygnus X North filament.

3. The atomic hydrogen producing the HISA absorption has a typical column density of $N$(H I) $\sim 1 \times 10^{20}$ cm$^{-2}$ in the DR21 filament. The timescale of H I-to-H$_2$ transition in the DR21 filament is calculated to be $3 \times 10^5$ yr, which approximates the ages of UCHII regions in the filament. This is in general consistent with a picture of rapid and dynamic cloud evolution.

We would like to thank the FAST staff for their support during the observations. This work is supported by National Key R&D Program of China No. 2022YFA1603100, No. 2017YFA0402604, and the National Natural Science Foundation of China (NSFC) grant U1731237. K.Q. acknowledges the science research grant from the China Manned Space Project with No. CMS-CSST-2021-B06. C.L. acknowledges the supports by NSFC grant 12103025, China Postdoctoral Science Foundation No. 2021M691532, and Jiangsu Postdoctoral Research Funding Program No. 2021K179B. Y.C. is partially supported by the Scholarship No. 201906190105 of the China Scholarship Council and the Predoctoral Program of the Smithsonian Astrophysical Observatory (SAO). This work makes use of publicly released data from the Nobeyama 45 m Cygnus X CO Survey. The Nobeyama Radio Observatory is a branch of the National Astronomical Observatory of Japan (NAOJ), National Institutes of Natural Science (NINS).

## Appendix A
## Velocity Channel Maps of the H I and $^{12}$CO Emissions

From the velocity channel maps (Figure 8), there appears to be anticorrelation between the HI and $^{12}$CO emissions.

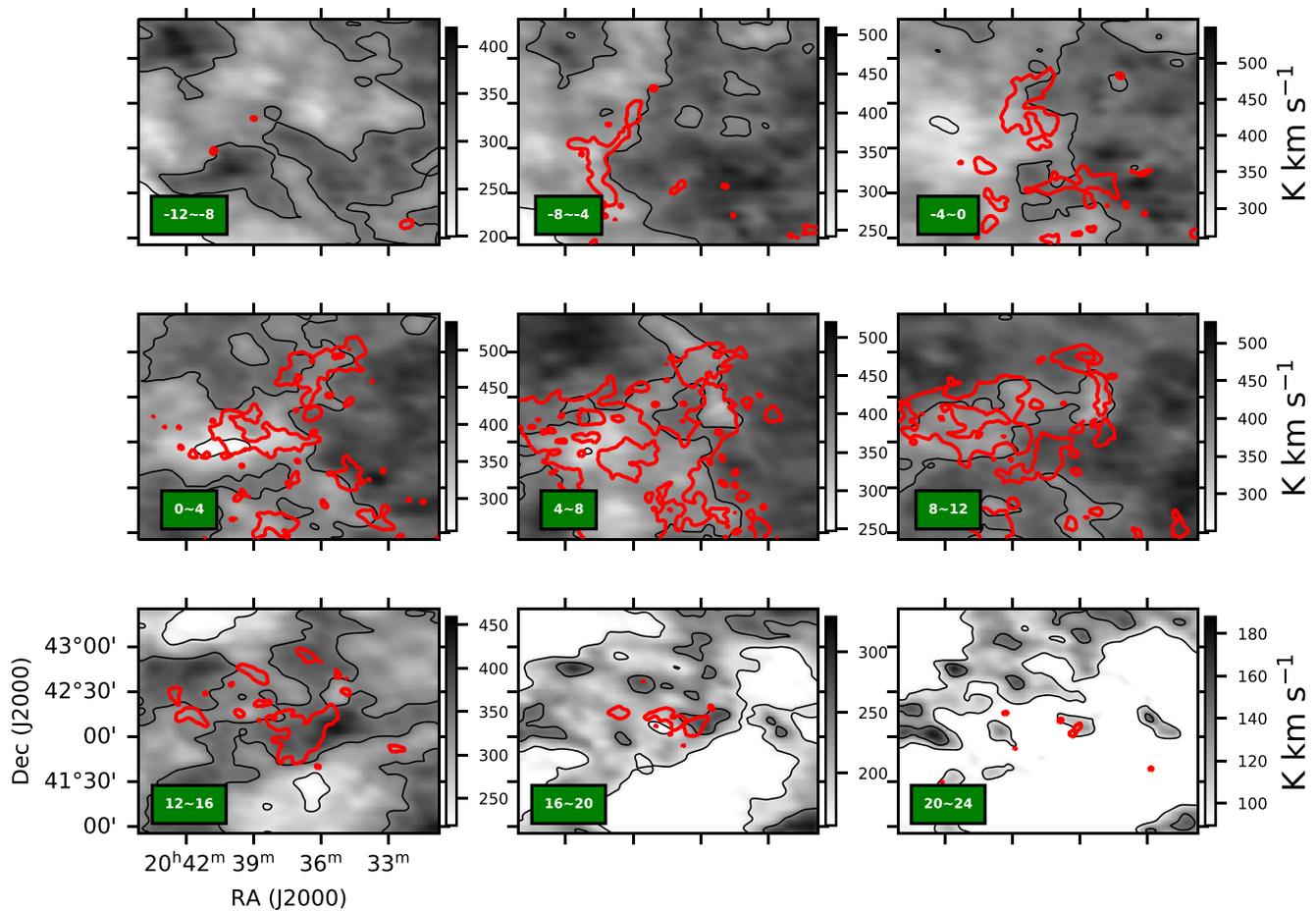

**Figure 8.** Velocity channel maps of H I (gray scale and gray contours) and $^{12}$CO (red contours) emissions between −12 and 24 km s$^{-1}$. The minimal level of the gray contours is 0.55 times the emission peak, and the interval is 0.25 times the peak. The minimal level of the red contours is 0.15 times the emission peak. Note that for display purposes, we have binned the original channels into velocity windows of a width of 4 km s$^{-1}$.

## Appendix B
## Comparison between the Line Center Velocities of the Extracted HISA Features and Observed $^{12}$CO Emission Lines

To extract the HISA spectra, we have tried the second-order, third-order, and fourth-order polynomial fittings and Gaussian fittings toward each pixel. To determine the best fitting results, we check the baselines of the HISA spectra (Figure 2) and also compare the line center velocities between the HISA and CO lines as shown in Figure 9.

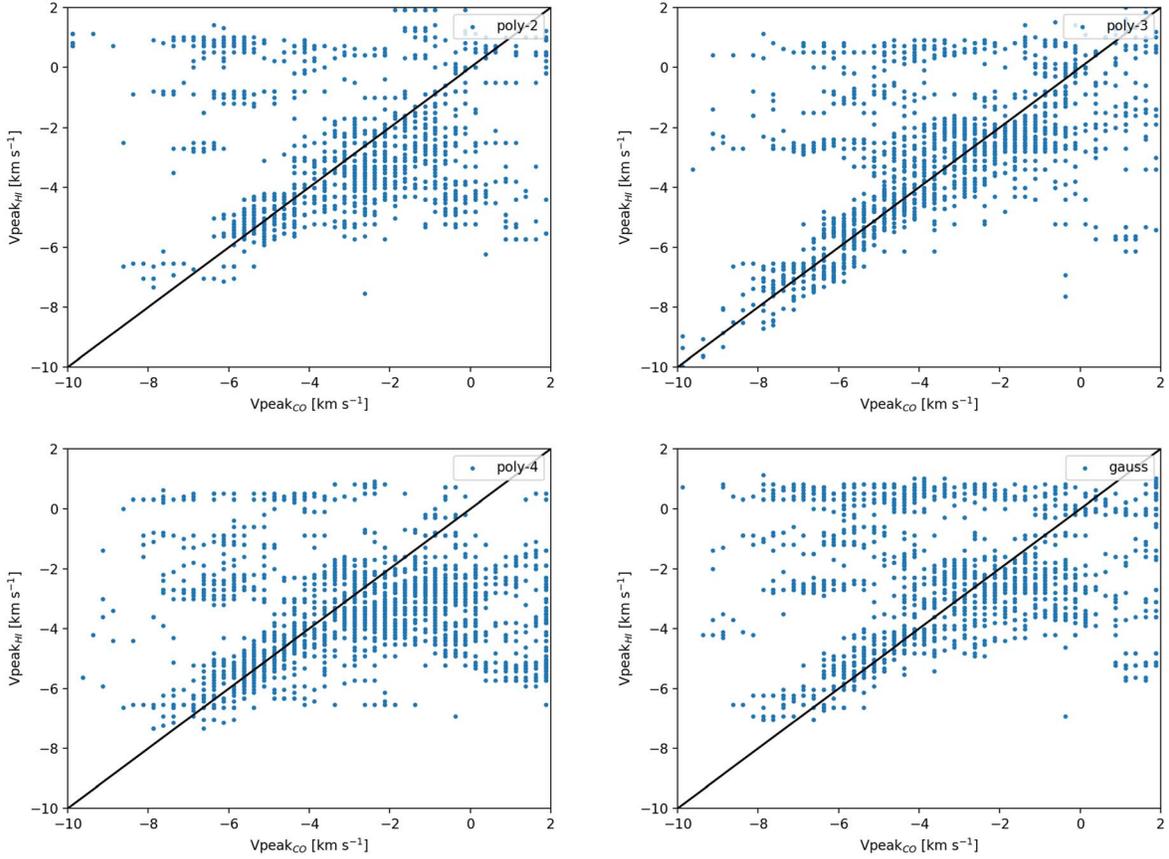

**Figure 9.** A pixel-by-pixel comparison between the line center velocity (relative to the local standard of rest) of the extracted HISA features vs. that of the observed $^{12}$CO emissions. As indicated in the top right corner of each panel, we show the comparisons for the HISA features derived with the second-order, third-order, and fourth-order polynomial fittings and with the Gaussian fittings. The solid line in each panel indicates the relation where the HISA velocity equals the $^{12}$CO velocity.

## Appendix C
## The Background Radio Continuum Brightness and the Derived HISA Optical Depth

From Equation (1), the background radio continuum brightness temperature, $T_{\rm cont}$, is needed to calculate the HISA optical depth. We take $T_{\rm cont}$ from the Effelsberg 21 cm radio continuum survey (Reich et al. 1990, 1997) as shown in the left panel of Figure 10 and adopt $p = 0.7$ and an excitation temperature of 40 K (see Section 3.1) to estimate the HISA optical depth. We also show the distribution of the derived optical depths at all the pixels in the right panel of Figure 10.

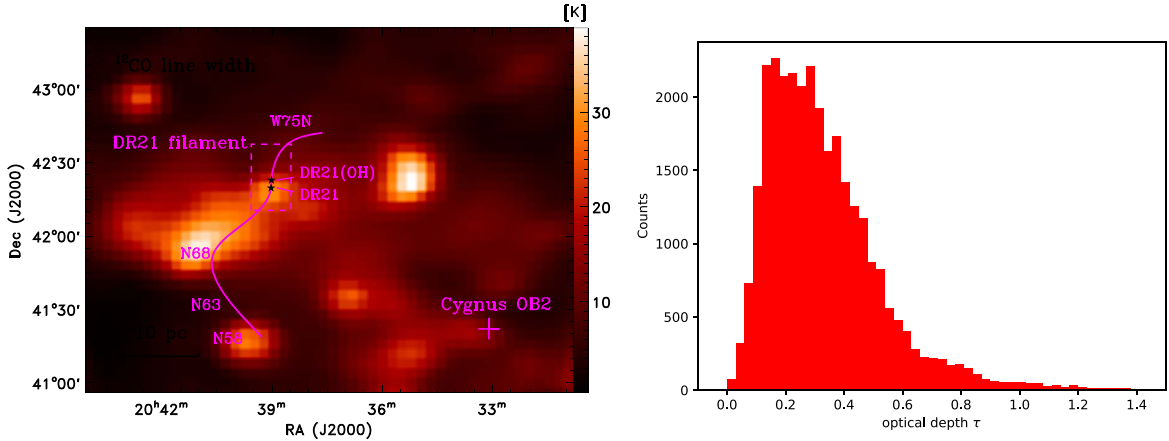

**Figure 10.** Left: $T_{\rm cont}$ from the Effelsberg 21 cm radio continuum emission. Right: counts histogram of the derived optical depth at all the pixels.

## ORCID iDs

Chong Li ⬤ https://orcid.org/0000-0003-2218-3437
Keping Qiu ⬤ https://orcid.org/0000-0002-5093-5088
Di Li ⬤ https://orcid.org/0000-0003-3010-7661
Hongchi Wang ⬤ https://orcid.org/0000-0003-0746-7968
Yue Cao ⬤ https://orcid.org/0000-0002-6368-7570
Junhao Liu ⬤ https://orcid.org/0000-0002-4774-2998
Yuehui Ma ⬤ https://orcid.org/0000-0002-8051-5228